# A Better Method for Volume Determination of Regularly and Irregularly Shaped Nanoparticles with Enhanced Accuracy

**Ravi Kiran Attota**
Nanoscale Metrology Group (NSMG)
Engineering Physics Division (EPD)
Physical Measurement Laboratory (PML)
National Institute of Standards and Technology (NIST)
Gaithersburg, MD 20899, USA
Ravikiran.attota@nist.gov; Ph: +1 301 975 5750

Nanoparticles (NPs) are widely used in diverse application areas, such as medicine, engineering, and cosmetics. The size (or volume) of NPs is one of the most important parameters for their successful application. It is relatively straightforward to determine the volume of regular NPs such as spheres and cubes from a one-dimensional or two-dimensional measurement. However, due to the three-dimensional nature of NPs, it is challenging to determine the proper physical size of many types of regularly and irregularly-shaped NPs (IS-NPs) at high-throughput using a single tool. Here, we present a relatively simple method that statistically determines a better volume estimate of many types of NPs by combining measurements from their top-down projection areas and peak-heights using two tools. The proposed method is significantly faster and more economical than the electron tomography method. We demonstrate the improved accuracy of the combined method over scanning electron microscopy (SEM) and atomic force microscopy (AFM) by using both modeling and measurements. This study also shows that SEM provides a more accurate estimate of size than AFM for most IS-NP size measurements. The method provides a much needed, proper high-throughput volumetric measurement method useful for many applications.

## Introduction

The projected market figure for nanotechnology incorporated in manufactured goods by 2020 is approximately 3,000 billion US dollars worldwide [1,2]. Nanomaterials in the form of nanoparticles (NPs) represent an important part of nanotechnology. NP usage is increasing at a fast pace with the development of new types of nanomaterials. It is reflected by the existence of over 1,300 diverse listed consumer goods [3] (e.g., sunscreens, cosmetics, clothes, foods, drugs, paints, varnishes, and self-cleaning coating for floors, walls and windows) in addition to medical [4] and high technology [5] applications. It has been reported that nanotechnology (through the use of NPs) has the potential to radically change the way cancer is treated [6]. Engineered NPs are a subset of nanomaterials, of which irregularly-shaped NPs (IS-NPs) are a major component. The environmental health and safety (EHS) of engineered NPs have attracted much attention owing to their potential toxicity. This is reflected by the amount of research (over 10,000 publications) that has been conducted on the subject of the environmental and health effects of nanomaterials in the last 15 years [7,8].

Several publications have highlighted the relationship between the size and toxicity of NPs [1,6-14]. They have also highlighted the lack of proper standards and characterization tools for NPs [1,6-8,10,15-17]. In fact, some have also complained that, “After over a decade of research, answers to the most basic questions are still lacking,” and suggest that more coherence in the experimental methods and materials used is needed [7,10]. One of the pressing issues with characterization is the large size discrepancy when the same set of NPs is measured using different tools [1,18-23]. Some publications have described variations in reported size even when NPs were measured using the same types of tool [1,19,20]. This creates confusion, as one of the fundamental starting points for application of NPs is an accurate knowledge of their size.

Electron tomography is a method that has demonstrated reliable 3D-shape reconstructions of nanoparticles[24-26]. But even though significant improvement has been made recently [27] , the technique is impractical and slow for routine and large-scale measurements required to statistically determine an accurate size and distribution of a batch of monodisperse NPs.

Apart from the electron tomography, at present, it is nearly impossible to accurately determine the size of many types of NPs(including types of IS-NPs[1,3]). For this reason, only reference materials (a less rigorous measurement certification process) exist for IS-NPs, such as gold [28,29]. However, for some nearly-spherical nanoparticles such as polystyrene, standard reference materials (a highly rigorous and accurate measurement process) exist [30]. Although several tools are currently available, more tools and methods are being sought [31,32], mainly due to a lack of consensus in the reported sizes. One of the latest additions to this long list of tools is through-focus scanning optical microscopy (TSOM)[33,34].

At this juncture, many critical questions remain: Is it possible to accurately measure the true volume of NPs, especially for IS-NPs and at high-throughput? If it is, which tool or method provides the correct spherical-volume-equivalent mean diameter at high throughput? If none, is there any way to determine which tool provides the smallest deviation from the accurate diameter? Is there any way to determine the deviation from the correct mean diameter so we can estimate the errors involved in the measured values? In this paper, we attempt to provide answers to these important fundamental questions.

## Background information

First, we briefly discuss the measurement procedures used to determine the reported diameters of NPs by the most widely used direct imaging tools, such as SEM, and AFM. These methods are usually considered to be the most accurate reference metrology tools, specifically when they are calibrated to an International System (SI) unit [18]. At the sub-100 nm scale, it is challenging to obtain complete 3-D information about IS-NPs. For this reason, certain assumptions are typically made when using these tools.

A two-dimensional, top-down area of projection is used to measure the diameters of IS-NPs using SEM (and also using conventional transmission electron microscopy (TEM)). Information regarding the height of the NPs is ignored. The calculated projected area equivalent diameter (Dxy) is reported as the SEM measured diameter (it is assumed that the height (or the 3$^{rd}$ dimension) is the same as the Dxy). In principle, both SEM and conventional TEM are supposed to produce the same size. However, in practice, SEM and TEM tools produce slightly different diameters due to differences in the edge detection methods.

Accurately calibrated AFM uses a sharp tip to measure the precise peak heights of NPs. However, it is extremely challenging to obtain an accurate contour (or area of projection) of NPs because of limitations in the physical interaction between the AFM tip and NPs. For this reason, only height information is used for the reported diameter and can be referred to as the peak-height equivalent diameter, which is the same as Dz. It is assumed that the AFM-measured height represents the diameter of perfectly spherical NPs and is reported as such.

In the case of SEM (and also for conventional TEM), only the 2-D projection area is exploited, whereas in the case of AFM, only the 1-D height information is exploited. Because none of the three tools use complete 3-D information on the IS-NPs, which are inherently 3-D in nature, they are prone to error. In fact, we should expect an erroneous diameter measurement and systematic difference in the reported diameters when SEM/TEM and AFM measurements are compared to each other. The published literature provides ample support for this observation (Table 1). We can also expect the reported diameters to include a certain deviation from the not yet known correct diameters. However, with the prevailing conventional knowledge, it is challenging to routinely determine the true diameter/volume of many types of NPs.

| Material | Mean diameters, nm | | | SEM Dia. | SEM Vol. | Ref. |
|---|---|---|---|---|---|---|
| | SEM | TEM | AFM | Devi. (%) | Devi. (%) | |
| SiO2 | 39.00 | 35.10 | 30.30 | 28.71 | 113.24 | [1] |
| SiO2 | 46.60 | 42.90 | 36.20 | 28.73 | 113.32 | [1] |
| SiO2 | 89.80 | 86.30 | 80.20 | 11.97 | 40.38 | [1] |
| Au | 85.48 | | 79.05 | 8.13 | 26.44 | [2] |
| Ag | 30.97 | | 23.14 | 33.84 | 139.74 | [11] |
| Au | 9.90 | | 7.20 | 37.50 | 159.96 | [16,17] |
| Au | 26.90 | | 23.70 | 13.50 | 46.22 | [16,17] |
| Au | 54.90 | | 53.90 | 1.86 | 5.67 | [16,17] |

**Table 1. Nanoparticle diameters reported in the literature.** A comparison of the reported mean diameters measured from the same batch using SEM, TEM and AFM shows disagreement in their values. SEM-reported diameters are always larger than the AFM-reported diameters. Diameters reported by SEM deviate upto 37 % from the diameters reported by AFM. Volumes calculated by SEM reported diameters deviate upto 150% from the volumes calculated by AFM reported diameters.

Ideally, we would like to have all types of NPs to be perfectly regular in nature such as spherical, cubic, or cylindrical. However, in reality, many types of NPs are irregular as shown in Fig. 1(a). In this study, we model these types of IS-NPs as ellipsoids (Figs. 1(d) to (d)).

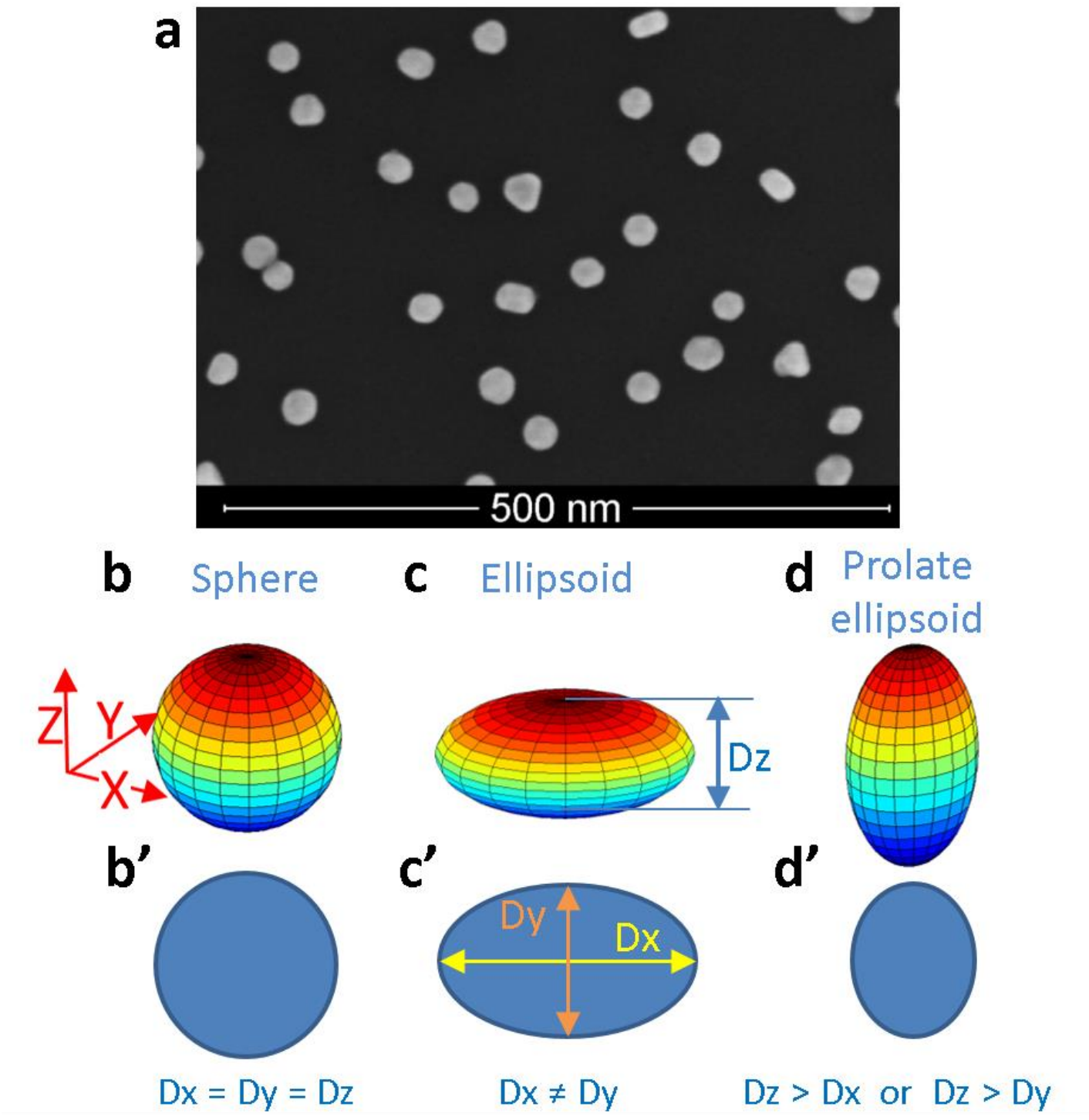

**Figure 1. Nanoparticle shapes.** (a) SEM image of nominally 30 nm diameter Au nanoparticles showing their irregular shapes. (b) to (d) Simplified ellipsoid shapes assigned to IS-NPs. (b') to (d'), Schematics of the corresponding areas of projections. Dx, Dy, and Dz are the diameters in the directions of the X, Y and Z axes, respectively.

Conventionally, the IS-NPs are first spread onto a substrate (or a grid) before measuring with SEM, TEM or AFM tools. This process typically aligns the IS-NPs in the most stable orientations on the substrates such that they have a wide base and a low center of gravity (or in this case due to van der Waals'/electrostatic force)[28]. Because of this, it is highly unlikely that the IS-NPs will come to rest in a prolate ellipsoid orientation (Fig. 1(d)), and hence we ignore this orientation in the following discussion.

## Results and discussion

Now, let us consider a 30 nm diameter spherical nanoparticle. This can be distorted into several ellipsoidal shapes while keeping the volume constant. The equivalent diameters that would be inferred for the different-shaped nanoparticles using SEM/TEM (Dxy) and AFM (Dz) are then evaluated as shown in Figures 2 (a) to (d) (the numerical values are presented in Table 2). Comparing the inferred Dxy with Dz, we can observe that the SEM measured diameters are always larger than the AFM measured diameters (except for a perfectly spherical particle). The same trend can be observed in the published literature (Table 1), indicating that the model we assume so far points in the right direction.

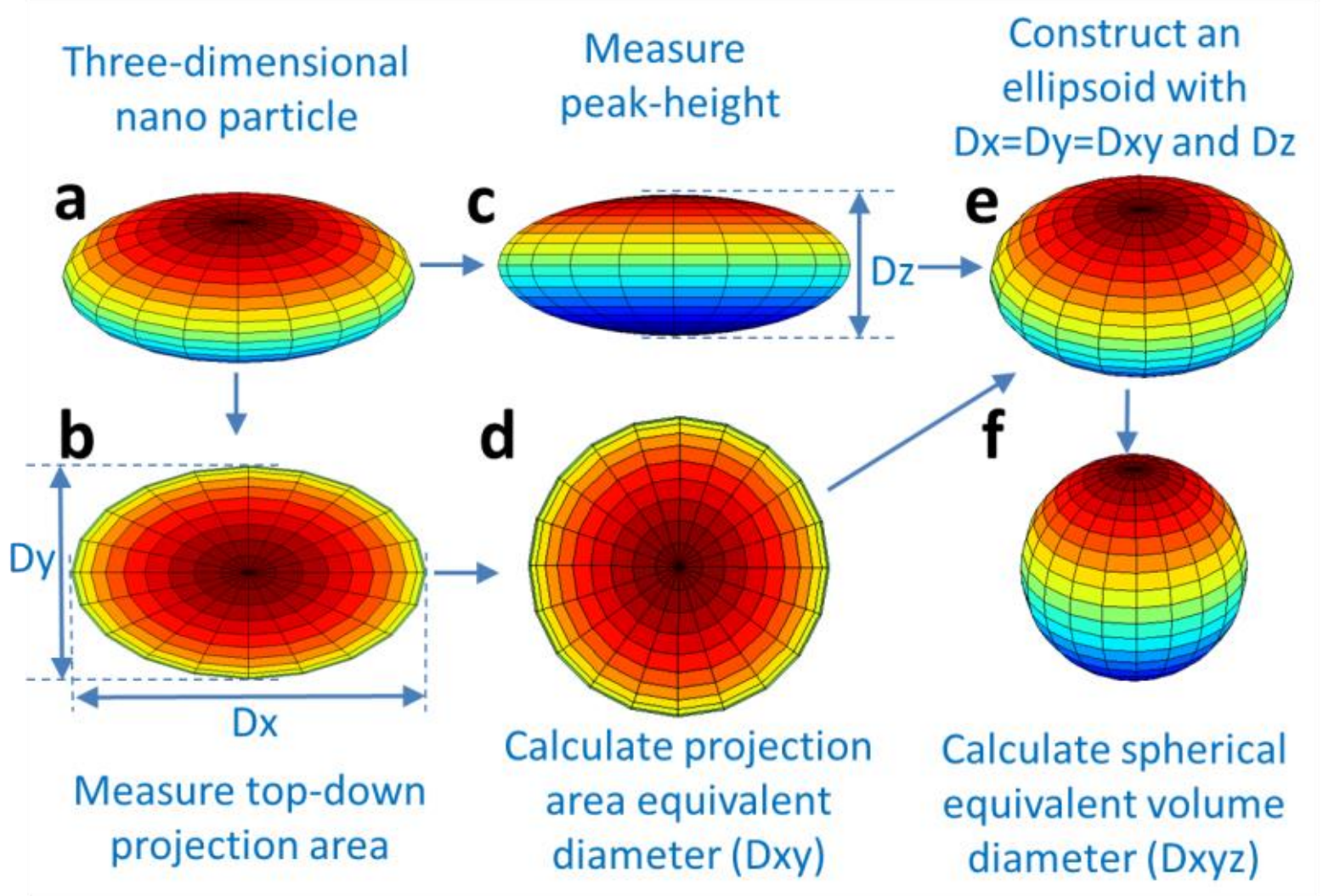

**Figure 2 Equivalent diameters calculation.** (a) A representative nanoparticle. This could be of any irregularly-shaped, three-dimensional nanoparticle. (b) Top-down area of projection as measured by tools such as SEM and TEM. For an IS-NP this would be an irregular area of projection. (c) Peak-height (Dz) as measured by tools such as AFM, and would be the reported diameter. (d) Calculated projection area equivalent diameter (Dxy) using the measured projection area from (b). Dxy would be the reported diameter using SEM or TEM. (e) Constructed ellipsoid by combining SEM/TEM and AFM reported diameters, where Dx = Dy = Dxy, and Dz = peak-height. (f) Calculated spherical equivalent volume diameter (Dxyz) from the constructed ellipsoid, where Dx = Dy = Dz = Dxyz.

Here we propose a third method to calculate the volume of some types of NPs, especially that of the type of IS-NPs shown in Fig. 1(a). In this method, top-down projection area as measured by SEM/TEM, and peak-height as measured by AFM are combined to obtain a 3D volume from which spherical equivalent volume diameter (Dxyz) is calculated as depicted in Figs. 2(e) and 2(f). Comparing the calculated Dz, Dxy and Dxyz in Table 2, we can see that the diameters determined by SEM/TEM are always larger while the diameters determined by AFM are always smaller than the correct diameter of 30 nm (except for the perfect spherical particle), which exactly equals Dxyz. We can also infer another minor point: the difference in the SEM/TEM and AFM measurements is proportional to the deviation of a NP from the perfect spherical shape.

| Dx | Dy | Dz | SEM Dia. (Dxy) | AFM Dia. (Dz) | Dxyz | SEM (Dxy) Dev. | AFM (Dz) Dev. | Shape ratios | | | |
|---|---|---|---|---|---|---|---|---|---|---|---|
| | | | | | | | | R1 | R2 | R3 | R4 |
| nm | nm | nm | nm | nm | nm | % | % | Dx/Dy | Dxy/Dz | R2/R1 | |
| 30.00 | 30.00 | 30.00 | 30.00 | 30.00 | 30.00 | 0.00 | 0.00 | 1.000 | 1.000 | 1.000 | 1.000 |
| 31.00 | 29.50 | 29.52 | 30.24 | 29.52 | 30.00 | 0.80 | -1.59 | 1.051 | 1.024 | 0.975 | -1.976 |
| 36.00 | 27.00 | 27.78 | 31.18 | 27.78 | 30.00 | 3.92 | -7.41 | 1.333 | 1.122 | 0.842 | -1.888 |
| 50.00 | 20.00 | 27.00 | 31.62 | 27.00 | 30.00 | 5.41 | -10.00 | 2.500 | 1.171 | 0.468 | -1.849 |
| 31.00 | 31.00 | 28.10 | 31.00 | 28.10 | 30.00 | 3.33 | -6.35 | 1.000 | 1.103 | 1.103 | -1.904 |
| 35.00 | 35.00 | 22.04 | 35.00 | 22.04 | 30.00 | 16.67 | -26.53 | 1.000 | 1.588 | 1.588 | -1.592 |
| 37.00 | 32.00 | 22.80 | 34.41 | 22.80 | 30.00 | 14.70 | -23.99 | 1.156 | 1.509 | 1.305 | -1.632 |

**Table 2. Inferred equivalent diameter evaluation.** A 30 nm diameter nanoparticle was distorted by keeping the volume constant to produce different Dx, Dy, and Dz diameters. Inferred equivalent diameters of the distorted nanoparticles were then calculated (Figure 2) as if they were measured using SEM/TEM (using the projection area), AFM (using the peak-height), and the combined method (using projection area and peak-height) proposed here. Dz = diameter in the Z-direction as measured by AFM, Dxy = projection area equivalent diameter as measured by SEM/TEM, Dxyz = spherical equivalent volume diameter determined by combining Dxy and Dz. If the shape ratio R3 is less than one, then the nanoparticles are dominated more by elongation than by flattening. If the shape ratio R3 is more than one, then the nanoparticles are dominated more by flattening than by elongation. R4 = (DzDev./Dxydev) or (AFMDev./SEMDev.)

In this section, we employ a similar analysis concept, but use experimentally measured realistic shape variations for Au nanoparticles, as an example. The shape variations thus obtained are then applied to 10 nm, 30 nm, 60 nm and 100 nm diameter nanoparticles for which the diameters as measured by SEM/TEM and AFM were calculated. Detailed results are presented in Tables S1(a) to S1(d). A summary of the results is presented in Table 3, from which the following important inferences can be made.

I. It appears possible to accurately obtain the spherical equivalent volume diameter of IS-NPs by combining SEM/TEM (projection-area) and AFM (peak-height) measurements.
II. The SEM/TEM-measured mean diameters are always larger than the AFM-measured mean diameters.
III. The SEM/TEM-measured diameters are larger, while the AFM-measured diameters are smaller than the correct equivalent diameters.

IV. The SEM/TEM measured diameters are more accurate than the AFM-measured diameters.

Nevertheless, the data presented in Tables 2 and 3 using partly modeled and partly experimental analysis cannot fully prove the accuracy of the proposed method; validation using experimental data is needed. Figure 15 in Ref. [3] provides this experimental data where for the first time, direct SEM and AFM measurements were provided for the same set of nanoparticles; suitable for the current type of analysis. The extracted SEM- and AFM-measured diameters and the calculated spherical equivalent volume diameters for each pair are tabulated in Supplementary Table S2. The average values obtained are presented in Table 4. AFM measured diameter is 5.48 nm smaller, while SEM measured diameter is 2.96 nm larger than the Dxyz.

| Dxyz | SEM (Dxy) | AFM (Dz) | SEM Dev. | AFM Dev. | Ratio |
|---|---|---|---|---|---|
| nm | nm | nm | % | % | R4 |
| 10.00 | 10.37 | 9.29 | 3.75 | -7.06 | -1.884 |
| 30.00 | 31.13 | 27.89 | 3.75 | -7.03 | -1.874 |
| 60.00 | 61.62 | 56.93 | 2.69 | -5.12 | -1.901 |
| 100.00 | 102.38 | 95.44 | 2.38 | -4.56 | -1.917 |

**Table 3. Calculated spherical equivalent volume diameters (Dxyz) using realistic shape variations of IS-NPs made of Au.** A summary of statistically derived Dxyz by combining the calculated SEM/TEM (Dxy) and AFM (Dz) diameters (from Supplementary Table S1) after applying measured shape variations of Au NPs to the four diameter sizes. SEM Dev. and AFM Dev. are the percentage deviations from the inferred correct diameters (Dxyz).

| Method | SEM (Dxy) | AFM (Dz) | Dxyz | SEM Dev. | AFM Dev. | Ratio |
|---|---|---|---|---|---|---|
| | nm | nm | nm | % | % | R4 |
| Direct | 58.17 | 49.73 | 55.21 | 5.36 | -9.92 | -1.85 |
| Adjusted | 55.00 | 49.73 | 53.18 | 3.41 | -6.49 | -1.90 |

**Table 4. A summary of the SEM and AFM data presented in Supplementary Table S2.** Dxyz in this table was calculated by combining the reported SEM and AFM diameters. Deviations of the reported SEM and AFM diameters from Dxyz were evaluated as a percentage. In the 'Adjusted' row, the SEM value was floated by maintaining a fixed AFM value such that the 'Ratio' magnitude equals 1.90.

The above shown experimental data are strikingly similar to those in Table 3. Identical inferences can be made using both the tables. The deviation ratios are also similar. This experimental data supports the modeled approach followed earlier in the paper and hence appears to confirm all the important inferences made above. However, we have not yet provided independent evidence for the accuracy of the proposed Dxyz. We provide this proof below.

Here we apply the combined method to evaluate the spherical equivalent volume diameter to irregularly-shaped macroparticles, such as pebbles. A key benefit of this approach is that the spherical equivalent volume diameter obtained using the projection area and the height can be corroborated by measuring the actual volume. This provides a definitive validation of the combined method presented in this paper. To perform this, we selected irregularly shaped aquarium glass pebbles weighing approximately 15 g to 21 g (Fig. 3). After following the procedure described above, we obtained a mean Dxyz of 23.8 mm using the projection area and the height of the 54 pebbles (Supplementary Table S3). The independently calculated Dxyz using the measured volume (obtained from mass and density) is 23.96 mm. As we can see, both the diameter values are statistically the same (from the statistical analysis given in Table S3) and hence confirm the accuracy of the combined method presented earlier. To obtain the true mean size, statistically-sufficient large number of particles must be measured. Of course, the particle size could range from nanoparticles to macroparticles. However, the 'Ratio' seems to depend on the nature and extent of the particle irregularities. In the case of gold NPs, the 'Ratio' varied from 1.88 to 1.92, depending on the size. However, in the case of pebbles, the measured 'Ratio' is approximately 1.63.

Since we have confirmed the modeled approach of the combined measurement presented earlier (Tables 2 and 3) with the experimental values (Supplementary Tables S2 and S3), we can now say with confidence that the most accurate Dxyz from the Au nanoparticle data presented in Ref. [3] is 55.21 nm (Supplementary Table S2). The SEM and AFM measured diameters are approximately 5.36 % and 9.92 % larger and smaller, respectively, than the Dxyz.

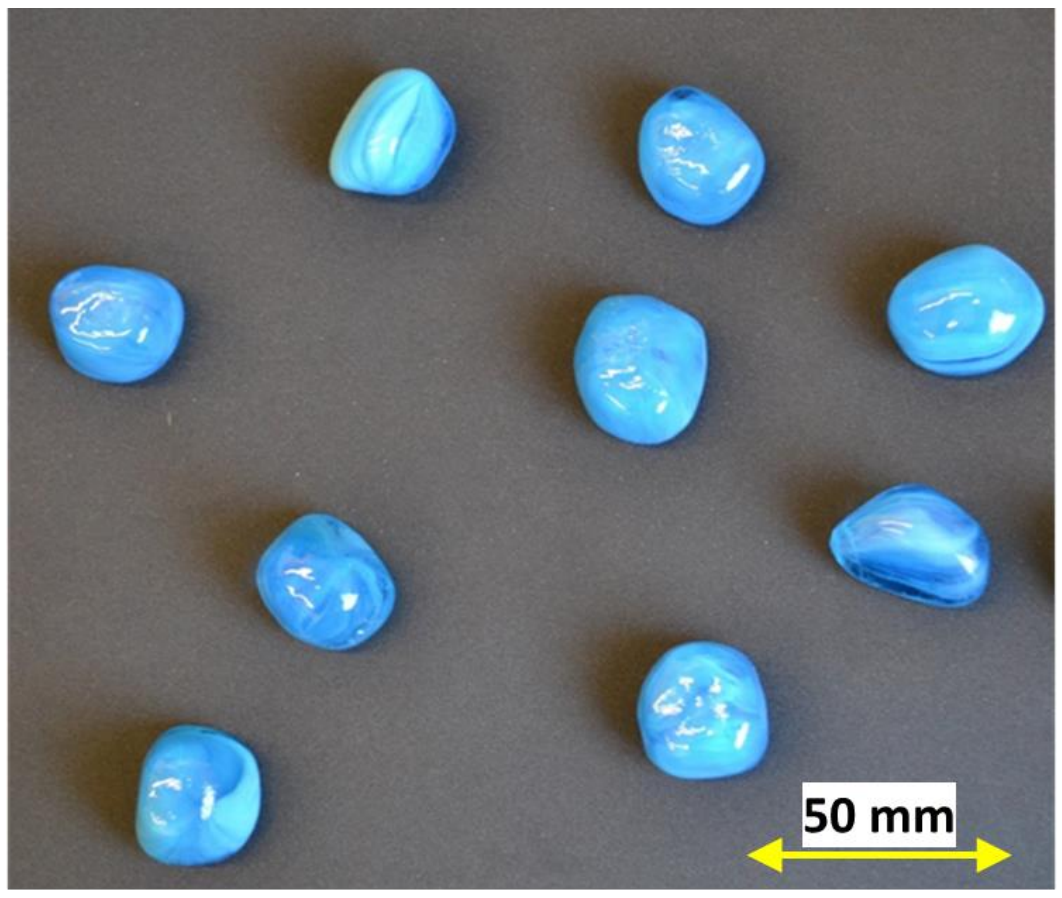

**Figure 3.** Macroscale pebbles used to compare the spherical equivalent volume diameters calculated based on Dxy and Dz; and from mass and density.

There is some degree of uncertainty associated with the SEM data presented in Ref. [3]. This stems from the fact that no accurate edge detection method has been developed for SEM images of Au nanoparticles. Because of this, the reported diameters of Au NPs were extracted at the full-width-half-max of the SEM intensity profiles [3]. This selection is a convenience based on the general nature of SEM imaging, but is not based on a rigorous model for imaging of Au NPs. Hence, we can expect an unknown measurement bias in the SEM measured diameters. However, no such wiggle room exists for the AFM measured diameters (peak-heights) and hence these can be considered fixed. Based on this, we can adjust the value of the SEM measured diameter such that the deviation ratio equals approximately 1.90 (from Table 3, based on the similarity in the size of particles to a 60 nm diameter and assuming similar shape irregularities). When the experimental data were adjusted to this ratio by keeping the AFM value constant and floating the SEM value, we obtain a new adjusted Dxyz of 53.18 nm (Table 4). The new adjusted SEM value is approximately 3.17 nm smaller than the reported value.

We can find some validation for this type of analysis from the published literature; however, we suggest caution. TEM measurements provide more accurate edge detection than SEM measurements, as it is possible to identify the end of atoms at the edges in TEM images more precisely than any non-model-based assignment of the physical edge location within the intensity profile of the SEM image. Because of this, we can expect a more accurate diameter measurement with TEM than with SEM when the measurements are performed on the same batch of NPs. The data presented in Table 1 show the measured diameters using SEM and TEM (in addition to AFM) for the same set of NPs. In these data, we can clearly see that the diameters measured using TEM are smaller than those measured using SEM, with a magnitude similar to the SEM adjusted value used in Table 4.

If the deviation in the diameter for a batch of monodispersed (in spherical equivalent volume diameters —or equivalently, volume) IS-NPs was known with certainty, then the measured diameters could be 'adjusted' by the known deviation to obtain the true diameter of Dxyz. For example, in Table 3 the AFM measured diameter (for 30 nm size) is 6.91 % smaller than the Dxyz. By adjusting the AFM measured diameter with this known deviation one could, in principle, obtain the Dxyz by using only one tool. This paper provides a basis for this type of adjustment.

For regularly-shaped, symmetric NPs such as spheres and cubes, the proposed combined method may not be of much use, as either SEM or AFM can provide a one dimensional measurement which will be same in other directions also. For perfect cylindrical-shaped NPs (with the axis along the substrate), SEM alone can provide all the necessary dimensions to calculate the volume; however, AFM cannot provide both necessary dimensions. For some other types of regularly-shaped NPs such as prisms, cuboids, and cylinders (resting on the circular end), the height information cannot be automatically obtained using SEM measurements alone, and lateral dimensions cannot be obtained using AFM alone. In such cases, again, combining SEM and AFM measurements will provide a more accurate volume determination compared to using either SEM or AFM alone; however, for these regular shapes, it is unnecessary and in fact preferable not to assume a spheroidal shape. Because of their regular shapes, basic geometric formulas can be used to determine their volumes from the two or three dimensions measured using SEM and/or AFM.

## Conclusions

In this paper, we have described a method to determine the volume of certain types of irregularly shaped, and regularly-shaped NPs by combining measurements obtained using top-down projection areas (as measured by tools such as SEM/TEM), and peak-heights (as measured by tools such as AFM) with more accuracy than can be measured by using either SEM or AFM alone. Both modeled and experimental evidence are provided to validate this concept. Based on the more accurate volumetric measurement of certain types of IS-NPs that this method provides, we demonstrate that for these IS-NPs, SEM/TEM measurements are more accurate than AFM measurements. Other NP measurement tools can also be evaluated for their accuracy. We expect this method to facilitate the certification of accurate reference materials for IS-NPs, which will increase the reliability and enhance trust in the results of studies of size-dependent properties of NPs, such as in toxicology.

**Acknowledgements**
The author would like to thank John Kramar for the useful discussions, and Andras Vladar for providing high-quality SEM images of Au nanoparticles used in Ref. [28].

**Competing financial interests**
The author declares no competing financial interests.

Supplementary information for

# A Better Method for Volume Determination of Regularly and Irregularly Shaped Nanoparticles with Enhanced Accuracy

Ravi Kiran Attota
NSMG, EPD, PML, NIST, Gaithersburg, MD 20899, USA
Ravikiran.attota@nist.gov

**Table S1. Spherical equivalent volume diameters (Dxyz) evaluated using realistic shape variations from measurements of Au nanoparticles.**

In Table 2, Dx and Dy values were randomly selected to demonstrate that combining SEM/TEM and AFM measurements will result in a correct Dxyz. Here we obtain realistic variations in Dx and Dy from the measured projection areas using actual data sets. For this we analyzed high-quality SEM images of nominally 10 nm, 30 nm, 60 nm and 100 nm diameter, irregularly shaped Au nanoparticles (RM certified by NIST [16,17]) using ImageJ to obtain areas of projection (Aproj), Dx (Feret maximum diameter), and Dy (Feret minimum diameter) for isolated nanoparticles. The Dx and Dy thus obtained are assumed to be the major and the minor axes, respectively, of the ellipsoid-shaped projected areas, even though they are not always exactly perpendicular to each other. With these assumptions we obtain typical, realistic shape variations from the projected areas. The measured shape variations thus obtained are then applied to (a) 10 nm, (b) 30 nm, (c) 60 nm, and (d) 100 nm diameter nanoparticles, and the diameters as measured by SEM/TEM and AFM are calculated.

Aproj = measured area of projection; Dx = measured maximum Feret diameter; Dy = measured minimum Feret diameter; Dxy = calculated diameter of a circle that has the same Aproj; $\Delta$Dx = percentage deviation in Dx from Dxy; $\Delta$Dy = percentage deviation in Dy from Dxy; Db = base diameter; Since this is a modeled calculation, Dxyz must be equal to Db.

By first evaluating $\Delta$Dx and $\Delta$Dy and then applying them to modeled sizes (Db), size variations (typically Gaussian distribution) present in the measured data (in addition to shape/area variations) within a size group was normalized.

Table S1 a

| S.No. | Measured data for nominally 10 nm diameter | | | | | | Measured ΔDx & ΔDy applied for Db = 100 nm | | | | | Shape ratios | | |
|---|---|---|---|---|---|---|---|---|---|---|---|---|---|---|
| | Aproj | Dx | Dy | Dxy | ΔDx | ΔDy | Dx | Dy | Dz(AFM) | Dxy(SEM) | Dxyz | R1 | R2 | R3 |
| | $nm^2$ | nm | nm | nm | % | % | nm | nm | nm | nm | nm | Dx/Dy | Dxy/Dz | R2/R1 |
| 1 | 138.52 | 14.89 | 12.77 | 13.28 | 12.13 | -3.86 | 11.21 | 9.61 | 9.28 | 10.38 | 10.00 | 1.166 | 1.119 | 0.960 |
| 2 | 126.71 | 14.34 | 12.13 | 12.70 | 12.95 | -4.50 | 11.30 | 9.55 | 9.27 | 10.39 | 10.00 | 1.183 | 1.120 | 0.947 |
| 3 | 109.19 | 12.93 | 11.42 | 11.79 | 9.64 | -3.12 | 10.96 | 9.69 | 9.41 | 10.31 | 10.00 | 1.132 | 1.095 | 0.967 |
| 4 | 105.52 | 13.02 | 10.85 | 11.59 | 12.34 | -6.37 | 11.23 | 9.36 | 9.51 | 10.26 | 10.00 | 1.200 | 1.079 | 0.899 |
| 5 | 114.89 | 13.33 | 11.86 | 12.09 | 10.22 | -1.89 | 11.02 | 9.81 | 9.25 | 10.40 | 10.00 | 1.123 | 1.124 | 1.001 |
| 6 | 123.04 | 14.70 | 11.49 | 12.51 | 17.43 | -8.19 | 11.74 | 9.18 | 9.28 | 10.38 | 10.00 | 1.279 | 1.119 | 0.875 |
| 7 | 123.04 | 13.65 | 12.64 | 12.51 | 9.04 | 0.99 | 10.90 | 10.10 | 9.08 | 10.49 | 10.00 | 1.080 | 1.156 | 1.070 |
| 8 | 105.12 | 12.93 | 10.85 | 11.57 | 11.75 | -6.19 | 11.17 | 9.38 | 9.54 | 10.24 | 10.00 | 1.191 | 1.073 | 0.901 |
| 9 | 96.15 | 12.93 | 10.21 | 11.06 | 16.84 | -7.68 | 11.68 | 9.23 | 9.27 | 10.39 | 10.00 | 1.266 | 1.120 | 0.885 |
| 10 | 96.97 | 12.33 | 10.85 | 11.11 | 10.97 | -2.32 | 11.10 | 9.77 | 9.23 | 10.41 | 10.00 | 1.136 | 1.129 | 0.993 |
| 11 | 112.04 | 12.93 | 11.49 | 11.94 | 8.24 | -3.79 | 10.82 | 9.62 | 9.60 | 10.20 | 10.00 | 1.125 | 1.063 | 0.945 |
| 12 | 110.01 | 13.66 | 11.69 | 11.83 | 15.45 | -1.23 | 11.54 | 9.88 | 8.77 | 10.68 | 10.00 | 1.169 | 1.218 | 1.042 |
| 13 | 122.64 | 13.57 | 12.13 | 12.49 | 8.62 | -2.92 | 10.86 | 9.71 | 9.48 | 10.27 | 10.00 | 1.119 | 1.083 | 0.968 |
| 14 | 112.04 | 13.42 | 11.49 | 11.94 | 12.37 | -3.79 | 11.24 | 9.62 | 9.25 | 10.40 | 10.00 | 1.168 | 1.124 | 0.962 |
| 15 | 131.60 | 14.01 | 12.77 | 12.94 | 8.29 | -1.36 | 10.83 | 9.86 | 9.36 | 10.34 | 10.00 | 1.098 | 1.104 | 1.006 |
| 16 | 101.86 | 12.57 | 11.28 | 11.39 | 10.43 | -0.89 | 11.04 | 9.91 | 9.14 | 10.46 | 10.00 | 1.114 | 1.145 | 1.028 |
| 17 | 112.45 | 14.00 | 11.27 | 11.96 | 17.02 | -5.79 | 11.70 | 9.42 | 9.07 | 10.50 | 10.00 | 1.242 | 1.157 | 0.932 |
| 18 | 116.52 | 13.57 | 11.49 | 12.18 | 11.43 | -5.66 | 11.14 | 9.43 | 9.51 | 10.25 | 10.00 | 1.181 | 1.078 | 0.913 |
| | | | | | | | | | | | | | | |
| | | | | | | | | Average -> | 9.29 | 10.37 | 10.00 | 1.165 | 1.117 | 0.961 |
| | | | | | | | | St.Dev. -> | 0.20 | 0.12 | 0.00 | 0.056 | 0.038 | 0.055 |

Table S1 b.

| S.No. | Measured data for nominally 30 nm diameter | | | | | | Measured ΔDx & ΔDy applied for Db = 30 nm | | | | | Shape ratios | | |
|---|---|---|---|---|---|---|---|---|---|---|---|---|---|---|
| | Aproj | Dx | Dy | Dxy | ΔDx | ΔDy | Dx | Dy | Dz(AFM) | Dxy(SEM) | Dxyz | R1 | R2 | R3 |
| | $nm^2$ | nm | nm | nm | % | % | nm | nm | nm | nm | nm | Dx/Dy | Dxy/Dz | R2/R1 |
| 1 | 478.23 | 27.00 | 24.54 | 24.67 | 9.45 | -0.51 | 32.83 | 29.85 | 27.55 | 31.30 | 30.00 | 1.100 | 1.136 | 1.033 |
| 2 | 531.73 | 27.85 | 25.86 | 26.01 | 7.07 | -0.59 | 32.12 | 29.82 | 28.18 | 30.95 | 30.00 | 1.077 | 1.098 | 1.020 |
| 3 | 541.77 | 29.85 | 25.25 | 26.26 | 13.69 | -3.85 | 34.11 | 28.85 | 27.44 | 31.37 | 30.00 | 1.182 | 1.143 | 0.967 |
| 4 | 603.63 | 29.49 | 27.99 | 27.72 | 6.38 | 0.98 | 31.92 | 30.29 | 27.93 | 31.09 | 30.00 | 1.054 | 1.113 | 1.057 |
| 5 | 494.95 | 27.28 | 25.30 | 25.10 | 8.68 | 0.80 | 32.61 | 30.24 | 27.38 | 31.40 | 30.00 | 1.078 | 1.147 | 1.063 |
| 6 | 688.91 | 34.43 | 27.76 | 29.61 | 16.28 | -6.24 | 34.88 | 28.13 | 27.52 | 31.33 | 30.00 | 1.240 | 1.138 | 0.918 |
| 7 | 551.80 | 29.32 | 26.52 | 26.50 | 10.63 | 0.06 | 33.19 | 30.02 | 27.10 | 31.56 | 30.00 | 1.106 | 1.165 | 1.053 |
| 8 | 595.27 | 32.33 | 24.57 | 27.52 | 17.45 | -10.74 | 35.24 | 26.78 | 28.62 | 30.72 | 30.00 | 1.316 | 1.073 | 0.816 |
| 9 | 707.31 | 37.05 | 25.41 | 30.00 | 23.49 | -15.31 | 37.05 | 25.41 | 28.69 | 30.68 | 30.00 | 1.458 | 1.070 | 0.733 |
| 10 | 499.96 | 28.36 | 21.98 | 25.23 | 12.43 | -12.85 | 33.73 | 26.14 | 30.62 | 29.69 | 30.00 | 1.290 | 0.970 | 0.752 |
| 11 | 551.80 | 30.08 | 25.86 | 26.50 | 13.50 | -2.41 | 34.05 | 29.28 | 27.09 | 31.57 | 30.00 | 1.163 | 1.166 | 1.002 |
| 12 | 715.67 | 32.41 | 28.45 | 30.18 | 7.37 | -5.74 | 32.21 | 28.28 | 29.64 | 30.18 | 30.00 | 1.139 | 1.018 | 0.894 |
| 13 | 571.86 | 32.33 | 23.72 | 26.98 | 19.83 | -12.08 | 35.95 | 26.38 | 28.48 | 30.79 | 30.00 | 1.363 | 1.081 | 0.793 |
| 14 | 553.47 | 30.74 | 24.57 | 26.54 | 15.81 | -7.43 | 34.74 | 27.77 | 27.98 | 31.06 | 30.00 | 1.251 | 1.110 | 0.887 |
| 15 | 580.22 | 31.09 | 25.86 | 27.17 | 14.40 | -4.83 | 34.32 | 28.55 | 27.55 | 31.30 | 30.00 | 1.202 | 1.136 | 0.945 |
| 16 | 792.58 | 35.51 | 29.74 | 31.76 | 11.80 | -6.36 | 33.54 | 28.09 | 28.66 | 30.69 | 30.00 | 1.194 | 1.071 | 0.897 |
| 17 | 474.88 | 27.00 | 24.57 | 24.58 | 9.83 | -0.06 | 32.95 | 29.98 | 27.33 | 31.43 | 30.00 | 1.099 | 1.150 | 1.046 |
| 18 | 504.98 | 28.39 | 25.45 | 25.35 | 11.98 | 0.40 | 33.59 | 30.12 | 26.68 | 31.81 | 30.00 | 1.115 | 1.192 | 1.069 |
| 19 | 473.21 | 27.49 | 21.98 | 24.54 | 12.02 | -10.42 | 33.61 | 26.87 | 29.90 | 30.05 | 30.00 | 1.251 | 1.005 | 0.804 |
| 20 | 496.62 | 27.91 | 23.28 | 25.14 | 11.03 | -7.42 | 33.31 | 27.77 | 29.18 | 30.42 | 30.00 | 1.199 | 1.042 | 0.869 |
| 21 | 469.87 | 27.67 | 23.95 | 24.45 | 13.17 | -2.05 | 33.95 | 29.38 | 27.07 | 31.58 | 30.00 | 1.155 | 1.167 | 1.010 |
| 22 | 511.67 | 28.39 | 24.57 | 25.52 | 11.25 | -3.72 | 33.37 | 28.88 | 28.01 | 31.05 | 30.00 | 1.155 | 1.108 | 0.959 |
| 23 | 700.62 | 34.14 | 28.45 | 29.86 | 14.33 | -4.73 | 34.30 | 28.58 | 27.54 | 31.31 | 30.00 | 1.200 | 1.137 | 0.947 |
| 24 | 570.19 | 29.77 | 25.86 | 26.94 | 10.51 | -4.00 | 33.15 | 28.80 | 28.28 | 30.90 | 30.00 | 1.151 | 1.093 | 0.949 |
| 25 | 570.19 | 29.54 | 25.86 | 26.94 | 9.67 | -4.00 | 32.90 | 28.80 | 28.49 | 30.78 | 30.00 | 1.142 | 1.080 | 0.946 |
| 26 | 449.80 | 27.19 | 23.28 | 23.93 | 13.62 | -2.72 | 34.09 | 29.18 | 27.14 | 31.54 | 30.00 | 1.168 | 1.162 | 0.995 |
| 27 | 489.93 | 28.48 | 24.54 | 24.97 | 14.04 | -1.71 | 34.21 | 29.49 | 26.76 | 31.76 | 30.00 | 1.160 | 1.187 | 1.023 |
| 28 | 598.62 | 33.94 | 22.50 | 27.60 | 22.97 | -18.50 | 36.89 | 24.45 | 29.93 | 30.03 | 30.00 | 1.509 | 1.003 | 0.665 |
| 29 | 603.63 | 32.41 | 25.86 | 27.72 | 16.91 | -6.69 | 35.07 | 27.99 | 27.50 | 31.33 | 30.00 | 1.253 | 1.139 | 0.909 |
| 30 | 580.22 | 31.49 | 25.77 | 27.17 | 15.88 | -5.18 | 34.76 | 28.44 | 27.31 | 31.45 | 30.00 | 1.222 | 1.152 | 0.942 |
| 31 | 667.17 | 32.12 | 28.35 | 29.14 | 10.23 | -2.71 | 33.07 | 29.19 | 27.97 | 31.07 | 30.00 | 1.133 | 1.111 | 0.980 |
| 32 | 596.95 | 31.57 | 26.55 | 27.56 | 14.53 | -3.69 | 34.36 | 28.89 | 27.20 | 31.51 | 30.00 | 1.189 | 1.159 | 0.974 |
| 33 | 478.23 | 27.67 | 23.77 | 24.67 | 12.17 | -3.64 | 33.65 | 28.91 | 27.75 | 31.19 | 30.00 | 1.164 | 1.124 | 0.965 |
| 34 | 548.45 | 29.77 | 25.86 | 26.42 | 12.67 | -2.11 | 33.80 | 29.37 | 27.20 | 31.51 | 30.00 | 1.151 | 1.158 | 1.006 |
| 35 | 630.39 | 31.94 | 27.16 | 28.33 | 12.75 | -4.13 | 33.83 | 28.76 | 27.75 | 31.19 | 30.00 | 1.176 | 1.124 | 0.956 |
| 36 | 573.54 | 29.52 | 25.86 | 27.02 | 9.25 | -4.28 | 32.77 | 28.72 | 28.69 | 30.68 | 30.00 | 1.141 | 1.069 | 0.937 |
| 37 | 601.96 | 30.19 | 27.16 | 27.68 | 9.06 | -1.89 | 32.72 | 29.43 | 28.04 | 31.03 | 30.00 | 1.112 | 1.107 | 0.996 |
| 38 | 513.34 | 27.91 | 25.60 | 25.56 | 9.21 | 0.16 | 32.76 | 30.05 | 27.43 | 31.38 | 30.00 | 1.090 | 1.144 | 1.049 |
| 39 | 553.47 | 29.49 | 25.86 | 26.54 | 11.10 | -2.56 | 33.33 | 29.23 | 27.71 | 31.21 | 30.00 | 1.140 | 1.126 | 0.988 |
| 40 | 499.96 | 28.36 | 24.57 | 25.23 | 12.43 | -2.60 | 33.73 | 29.22 | 27.40 | 31.39 | 30.00 | 1.154 | 1.146 | 0.993 |
| 41 | 652.13 | 31.46 | 28.45 | 28.81 | 9.21 | -1.25 | 32.76 | 29.62 | 27.82 | 31.15 | 30.00 | 1.106 | 1.120 | 1.013 |
| 42 | 551.80 | 29.06 | 26.50 | 26.50 | 9.65 | 0.00 | 32.90 | 30.00 | 27.36 | 31.42 | 30.00 | 1.096 | 1.148 | 1.047 |
| 43 | 635.40 | 32.74 | 28.45 | 28.44 | 15.13 | 0.04 | 34.54 | 30.01 | 26.05 | 32.19 | 30.00 | 1.151 | 1.236 | 1.074 |
| 44 | 615.34 | 30.30 | 27.16 | 27.98 | 8.27 | -2.97 | 32.48 | 29.11 | 28.56 | 30.75 | 30.00 | 1.116 | 1.077 | 0.965 |
| 45 | 531.73 | 28.92 | 24.57 | 26.01 | 11.15 | -5.56 | 33.34 | 28.33 | 28.58 | 30.74 | 30.00 | 1.177 | 1.076 | 0.914 |
| 46 | 638.75 | 31.94 | 27.92 | 28.51 | 12.01 | -2.07 | 33.60 | 29.38 | 27.35 | 31.42 | 30.00 | 1.144 | 1.149 | 1.004 |
| 47 | 678.88 | 34.82 | 27.19 | 29.39 | 18.45 | -7.51 | 35.54 | 27.75 | 27.38 | 31.40 | 30.00 | 1.281 | 1.147 | 0.895 |
| 48 | 493.27 | 28.24 | 23.28 | 25.06 | 12.72 | -7.10 | 33.81 | 27.87 | 28.65 | 30.70 | 30.00 | 1.213 | 1.071 | 0.883 |
| 49 | 588.59 | 30.27 | 27.18 | 27.37 | 10.60 | -0.68 | 33.18 | 29.80 | 27.31 | 31.44 | 30.00 | 1.114 | 1.151 | 1.034 |
| 50 | 596.95 | 31.94 | 24.69 | 27.56 | 15.87 | -10.43 | 34.76 | 26.87 | 28.91 | 30.56 | 30.00 | 1.294 | 1.057 | 0.817 |
| 51 | 615.34 | 35.18 | 22.54 | 27.98 | 25.70 | -19.47 | 37.71 | 24.16 | 29.64 | 30.18 | 30.00 | 1.561 | 1.018 | 0.652 |
| 52 | 643.77 | 31.09 | 28.45 | 28.62 | 8.61 | -0.62 | 32.58 | 29.82 | 27.79 | 31.17 | 30.00 | 1.093 | 1.121 | 1.026 |
| 53 | 847.76 | 36.78 | 32.00 | 32.85 | 11.97 | -2.57 | 33.59 | 29.23 | 27.50 | 31.33 | 30.00 | 1.149 | 1.139 | 0.991 |
| 54 | 837.73 | 37.05 | 32.33 | 32.65 | 13.47 | -0.99 | 34.04 | 29.70 | 26.70 | 31.80 | 30.00 | 1.146 | 1.191 | 1.039 |
| 55 | 608.65 | 30.66 | 27.16 | 27.83 | 10.14 | -2.43 | 33.04 | 29.27 | 27.92 | 31.10 | 30.00 | 1.129 | 1.114 | 0.987 |
| 56 | 633.73 | 31.30 | 28.45 | 28.40 | 10.22 | 0.17 | 33.07 | 30.05 | 27.17 | 31.52 | 30.00 | 1.100 | 1.160 | 1.054 |
| 57 | 503.31 | 30.19 | 24.47 | 25.31 | 19.28 | -3.34 | 35.78 | 29.00 | 26.02 | 32.21 | 30.00 | 1.234 | 1.238 | 1.003 |
| 58 | 810.98 | 35.32 | 31.98 | 32.13 | 9.94 | -0.45 | 32.98 | 29.87 | 27.41 | 31.38 | 30.00 | 1.104 | 1.145 | 1.037 |
| 59 | 688.91 | 32.64 | 28.45 | 29.61 | 10.22 | -3.93 | 33.07 | 28.82 | 28.33 | 30.87 | 30.00 | 1.147 | 1.090 | 0.950 |
| 60 | 506.65 | 27.76 | 24.57 | 25.39 | 9.34 | -3.25 | 32.80 | 29.03 | 28.36 | 30.86 | 30.00 | 1.130 | 1.088 | 0.963 |
| | | | | | | | | | | | | | | |
| | | | | | | | | Average -> | 27.89 | 31.13 | 30.00 | 1.182 | 1.118 | 0.953 |
| | | | | | | | | St.Dev. -> | 0.91 | 0.50 | 0.00 | 0.099 | 0.053 | 0.096 |

Table S1 c.

| | Measured data for nominally 60 nm diameter | | | | | | Measured ΔDx & ΔDy applied for Db = 60 nm | | | | | Shape ratios | | |
|---|---|---|---|---|---|---|---|---|---|---|---|---|---|---|
| S.No. | Aproj | Dx | Dy | Dxy | ΔDx | ΔDy | Dx | Dy | Dz(AFM) | Dxy(SEM) | Dxyz | R1 | R2 | R3 |
| | $nm^2$ | nm | nm | nm | % | % | nm | nm | nm | nm | nm | Dx/Dy | Dxy/Dz | R2/R1 |
| 1 | 2150.47 | 57.45 | 51.65 | 52.32 | 9.82 | -1.27 | 65.89 | 59.24 | 55.34 | 62.47 | 60.00 | 1.112 | 1.129 | 1.015 |
| 2 | 2266.16 | 57.92 | 52.17 | 53.70 | 7.85 | -2.85 | 64.71 | 58.29 | 57.26 | 61.42 | 60.00 | 1.110 | 1.073 | 0.966 |
| 3 | 2027.98 | 58.62 | 48.72 | 50.80 | 15.39 | -4.10 | 69.23 | 57.54 | 54.22 | 63.12 | 60.00 | 1.203 | 1.164 | 0.967 |
| 4 | 2395.46 | 60.23 | 52.17 | 55.22 | 9.07 | -5.51 | 65.44 | 56.69 | 58.22 | 60.91 | 60.00 | 1.154 | 1.046 | 0.906 |
| 5 | 2837.81 | 74.75 | 49.81 | 60.10 | 24.38 | -17.13 | 74.63 | 49.72 | 58.21 | 60.92 | 60.00 | 1.501 | 1.046 | 0.697 |
| 6 | 2368.24 | 61.57 | 53.49 | 54.90 | 12.14 | -2.56 | 67.29 | 58.46 | 54.91 | 62.72 | 60.00 | 1.151 | 1.142 | 0.992 |
| 7 | 2204.92 | 58.33 | 52.00 | 52.97 | 10.11 | -1.84 | 66.07 | 58.89 | 55.51 | 62.38 | 60.00 | 1.122 | 1.124 | 1.002 |
| 8 | 2504.35 | 65.69 | 49.57 | 56.46 | 16.35 | -12.21 | 69.81 | 52.68 | 58.74 | 60.64 | 60.00 | 1.325 | 1.032 | 0.779 |
| 9 | 2361.44 | 62.23 | 53.24 | 54.82 | 13.51 | -2.88 | 68.10 | 58.27 | 54.43 | 63.00 | 60.00 | 1.169 | 1.157 | 0.990 |
| 10 | 2259.36 | 61.07 | 49.33 | 53.62 | 13.88 | -8.02 | 68.33 | 55.19 | 57.28 | 61.41 | 60.00 | 1.238 | 1.072 | 0.866 |
| 11 | 1932.70 | 56.44 | 48.38 | 49.60 | 13.79 | -2.45 | 68.27 | 58.53 | 54.05 | 63.21 | 60.00 | 1.166 | 1.169 | 1.003 |
| 12 | 2170.89 | 56.19 | 52.17 | 52.56 | 6.90 | -0.74 | 64.14 | 59.56 | 56.54 | 61.81 | 60.00 | 1.077 | 1.093 | 1.015 |
| 13 | 2096.03 | 60.23 | 46.69 | 51.65 | 16.61 | -9.61 | 69.96 | 54.23 | 56.93 | 61.60 | 60.00 | 1.290 | 1.082 | 0.839 |
| 14 | 2041.59 | 55.77 | 49.57 | 50.97 | 9.40 | -2.76 | 65.64 | 58.34 | 56.40 | 61.88 | 60.00 | 1.125 | 1.097 | 0.975 |
| 15 | 2640.45 | 63.10 | 58.35 | 57.97 | 8.84 | 0.65 | 65.30 | 60.39 | 54.77 | 62.80 | 60.00 | 1.081 | 1.147 | 1.060 |
| 16 | 1851.04 | 52.17 | 46.96 | 48.54 | 7.49 | -3.26 | 64.50 | 58.05 | 57.70 | 61.19 | 60.00 | 1.111 | 1.060 | 0.954 |
| 17 | 2300.19 | 58.86 | 51.98 | 54.11 | 8.78 | -3.93 | 65.27 | 57.64 | 57.42 | 61.34 | 60.00 | 1.132 | 1.068 | 0.943 |
| 18 | 2204.92 | 59.49 | 52.17 | 52.97 | 12.29 | -1.51 | 67.38 | 59.09 | 54.25 | 63.10 | 60.00 | 1.140 | 1.163 | 1.020 |
| 19 | 1959.92 | 54.28 | 49.57 | 49.94 | 8.69 | -0.76 | 65.21 | 59.54 | 55.63 | 62.31 | 60.00 | 1.095 | 1.120 | 1.023 |
| 20 | 2300.19 | 65.22 | 46.36 | 54.11 | 20.53 | -14.32 | 72.32 | 51.41 | 58.10 | 60.97 | 60.00 | 1.407 | 1.049 | 0.746 |
| 21 | 2048.39 | 55.03 | 49.57 | 51.06 | 7.78 | -2.93 | 64.67 | 58.24 | 57.35 | 61.37 | 60.00 | 1.110 | 1.070 | 0.964 |
| 22 | 2238.94 | 59.49 | 52.17 | 53.38 | 11.44 | -2.26 | 66.86 | 58.64 | 55.09 | 62.62 | 60.00 | 1.140 | 1.137 | 0.997 |
| 23 | 1653.69 | 49.22 | 44.35 | 45.88 | 7.29 | -3.33 | 64.37 | 58.00 | 57.85 | 61.10 | 60.00 | 1.110 | 1.056 | 0.952 |
| 24 | 2116.45 | 57.27 | 51.98 | 51.90 | 10.35 | 0.15 | 66.21 | 60.09 | 54.29 | 63.08 | 60.00 | 1.102 | 1.162 | 1.055 |
| 25 | 1776.18 | 52.82 | 46.96 | 47.55 | 11.10 | -1.24 | 66.66 | 59.26 | 54.68 | 62.85 | 60.00 | 1.125 | 1.149 | 1.022 |
| 26 | 1980.34 | 55.03 | 49.57 | 50.20 | 9.61 | -1.27 | 65.77 | 59.24 | 55.44 | 62.42 | 60.00 | 1.110 | 1.126 | 1.014 |
| 27 | 2347.83 | 60.06 | 52.17 | 54.66 | 9.87 | -4.55 | 65.92 | 57.27 | 57.22 | 61.44 | 60.00 | 1.151 | 1.074 | 0.933 |
| 28 | 2136.86 | 58.62 | 51.26 | 52.15 | 12.41 | -1.72 | 67.45 | 58.97 | 54.31 | 63.07 | 60.00 | 1.144 | 1.161 | 1.015 |
| 29 | 1943.03 | 53.45 | 47.16 | 49.73 | 7.48 | -5.16 | 64.49 | 56.90 | 58.86 | 60.58 | 60.00 | 1.133 | 1.029 | 0.908 |
| 30 | 2429.61 | 58.74 | 53.69 | 55.61 | 5.63 | -3.45 | 63.38 | 57.93 | 58.83 | 60.59 | 60.00 | 1.094 | 1.030 | 0.942 |
| 31 | 2391.17 | 60.19 | 53.03 | 55.17 | 9.10 | -3.86 | 65.46 | 57.68 | 57.21 | 61.45 | 60.00 | 1.135 | 1.074 | 0.946 |
| 32 | 2127.56 | 56.44 | 50.39 | 52.04 | 8.46 | -3.16 | 65.08 | 58.10 | 57.12 | 61.49 | 60.00 | 1.120 | 1.076 | 0.961 |
| 33 | 1796.95 | 50.15 | 47.16 | 47.82 | 4.87 | -1.38 | 62.92 | 59.17 | 58.01 | 61.02 | 60.00 | 1.063 | 1.052 | 0.989 |
| 34 | 2351.63 | 58.66 | 51.01 | 54.71 | 7.23 | -6.76 | 64.34 | 55.95 | 60.01 | 59.99 | 60.00 | 1.150 | 1.000 | 0.869 |
| 35 | 2136.34 | 59.58 | 44.42 | 52.14 | 14.26 | -14.82 | 68.56 | 51.11 | 61.64 | 59.19 | 60.00 | 1.341 | 0.960 | 0.716 |
| 36 | 3002.96 | 68.96 | 61.12 | 61.82 | 11.55 | -1.13 | 66.93 | 59.32 | 54.40 | 63.01 | 60.00 | 1.128 | 1.158 | 1.027 |
| 37 | 1854.06 | 53.04 | 46.69 | 48.58 | 9.18 | -3.89 | 65.51 | 57.67 | 57.18 | 61.46 | 60.00 | 1.136 | 1.075 | 0.946 |
| 38 | 2275.84 | 56.41 | 53.72 | 53.82 | 4.81 | -0.19 | 62.89 | 59.89 | 57.35 | 61.37 | 60.00 | 1.050 | 1.070 | 1.019 |
| 39 | 2399.95 | 58.28 | 53.44 | 55.27 | 5.45 | -3.30 | 63.27 | 58.02 | 58.84 | 60.59 | 60.00 | 1.090 | 1.030 | 0.944 |
| 40 | 3121.59 | 65.59 | 62.72 | 63.03 | 4.06 | -0.49 | 62.44 | 59.70 | 57.94 | 61.06 | 60.00 | 1.046 | 1.054 | 1.008 |
| 41 | 2441.69 | 64.78 | 51.61 | 55.75 | 16.21 | -7.42 | 69.73 | 55.55 | 55.77 | 62.24 | 60.00 | 1.255 | 1.116 | 0.889 |
| 42 | 2156.11 | 57.56 | 51.09 | 52.38 | 9.87 | -2.47 | 65.92 | 58.52 | 55.99 | 62.11 | 60.00 | 1.126 | 1.109 | 0.985 |
| 43 | 2147.33 | 55.90 | 51.36 | 52.28 | 6.93 | -1.76 | 64.16 | 58.94 | 57.12 | 61.50 | 60.00 | 1.088 | 1.077 | 0.989 |
| 44 | 1972.69 | 53.87 | 48.78 | 50.11 | 7.51 | -2.65 | 64.51 | 58.41 | 57.33 | 61.38 | 60.00 | 1.104 | 1.071 | 0.969 |
| 45 | 2574.60 | 62.55 | 56.04 | 57.24 | 9.27 | -2.10 | 65.56 | 58.74 | 56.09 | 62.06 | 60.00 | 1.116 | 1.106 | 0.991 |
| 46 | 2160.51 | 56.63 | 50.31 | 52.44 | 8.00 | -4.07 | 64.80 | 57.56 | 57.91 | 61.07 | 60.00 | 1.126 | 1.055 | 0.937 |
| 47 | 2238.49 | 60.76 | 49.03 | 53.38 | 13.83 | -8.14 | 68.30 | 55.12 | 57.38 | 61.35 | 60.00 | 1.239 | 1.069 | 0.863 |
| 48 | 2453.78 | 61.77 | 55.02 | 55.88 | 10.54 | -1.54 | 66.32 | 59.07 | 55.13 | 62.59 | 60.00 | 1.123 | 1.135 | 1.011 |
| 49 | 2172.59 | 56.50 | 50.39 | 52.58 | 7.44 | -4.17 | 64.46 | 57.50 | 58.27 | 60.88 | 60.00 | 1.121 | 1.045 | 0.932 |
| 50 | 2069.34 | 55.56 | 49.38 | 51.32 | 8.25 | -3.77 | 64.95 | 57.74 | 57.60 | 61.24 | 60.00 | 1.125 | 1.063 | 0.945 |
| 51 | 2416.43 | 62.08 | 51.55 | 55.46 | 11.95 | -7.05 | 67.17 | 55.77 | 57.66 | 61.20 | 60.00 | 1.204 | 1.061 | 0.881 |
| 52 | 2198.95 | 56.53 | 51.35 | 52.90 | 6.85 | -2.93 | 64.11 | 58.24 | 57.85 | 61.11 | 60.00 | 1.101 | 1.056 | 0.960 |
| 53 | 2074.84 | 54.71 | 50.68 | 51.39 | 6.46 | -1.39 | 63.88 | 59.17 | 57.15 | 61.48 | 60.00 | 1.080 | 1.076 | 0.996 |
| 54 | 2438.40 | 60.80 | 51.39 | 55.71 | 9.13 | -7.76 | 65.48 | 55.34 | 59.61 | 60.20 | 60.00 | 1.183 | 1.010 | 0.854 |
| 55 | 2236.49 | 57.09 | 51.13 | 53.35 | 7.00 | -4.16 | 64.20 | 57.50 | 58.51 | 60.76 | 60.00 | 1.117 | 1.038 | 0.930 |
| 56 | 3116.28 | 68.03 | 61.64 | 62.98 | 8.02 | -2.13 | 64.81 | 58.72 | 56.75 | 61.69 | 60.00 | 1.104 | 1.087 | 0.985 |
| 57 | 2665.50 | 63.09 | 56.35 | 58.24 | 8.33 | -3.26 | 65.00 | 58.05 | 57.25 | 61.42 | 60.00 | 1.120 | 1.073 | 0.958 |
| 58 | 2172.25 | 55.74 | 51.98 | 52.58 | 6.00 | -1.14 | 63.60 | 59.32 | 57.26 | 61.42 | 60.00 | 1.072 | 1.073 | 1.000 |
| 59 | 2207.09 | 57.72 | 51.13 | 53.00 | 8.91 | -3.53 | 65.35 | 57.88 | 57.11 | 61.50 | 60.00 | 1.129 | 1.077 | 0.954 |
| 60 | 2259.36 | 57.77 | 51.19 | 53.62 | 7.73 | -4.54 | 64.64 | 57.28 | 58.34 | 60.85 | 60.00 | 1.129 | 1.043 | 0.924 |
| | | | | | | | | | | | | | | |
| | | | | | | | | Average -> | 56.93 | 61.62 | 60.00 | 1.148 | 1.084 | 0.949 |
| | | | | | | | | St.Dev. -> | 1.61 | 0.87 | 0.00 | 0.083 | 0.046 | 0.076 |

Table S1 d.

| S.No. | Measured data for nominally 100 nm diameter | | | | | | Measured ΔDx & ΔDy applied for Db = 100 nm | | | | | Shape ratios | | |
|---|---|---|---|---|---|---|---|---|---|---|---|---|---|---|
| | Aproj | Dx | Dy | Dxy | ΔDx | ΔDy | Dx | Dy | Dz(AFM) | Dxy(SEM) | Dxyz | R1 | R2 | R3 |
| | $nm^2$ | nm | nm | nm | % | % | nm | nm | nm | nm | nm | Dx/Dy | Dxy/Dz | R2/R1 |
| 1 | 7957.22 | 108.36 | 97.66 | 100.63 | 7.68 | -2.96 | 107.68 | 97.04 | 95.70 | 102.22 | 100.00 | 1.110 | 1.068 | 0.963 |
| 2 | 7480.38 | 104.50 | 96.10 | 97.57 | 7.10 | -1.51 | 107.10 | 98.49 | 94.81 | 102.70 | 100.00 | 1.087 | 1.083 | 0.996 |
| 3 | 8577.11 | 123.55 | 88.04 | 104.48 | 18.25 | -15.73 | 118.25 | 84.27 | 100.35 | 99.82 | 100.00 | 1.403 | 0.995 | 0.709 |
| 4 | 9584.43 | 116.47 | 109.86 | 110.45 | 5.46 | -0.53 | 105.46 | 99.47 | 95.33 | 102.42 | 100.00 | 1.060 | 1.074 | 1.013 |
| 5 | 8761.88 | 112.89 | 104.74 | 105.60 | 6.90 | -0.81 | 106.90 | 99.19 | 94.31 | 102.97 | 100.00 | 1.078 | 1.092 | 1.013 |
| 6 | 8589.03 | 112.86 | 102.54 | 104.55 | 7.95 | -1.93 | 107.95 | 98.07 | 94.46 | 102.89 | 100.00 | 1.101 | 1.089 | 0.990 |
| 7 | 7629.40 | 103.58 | 97.66 | 98.54 | 5.11 | -0.90 | 105.11 | 99.10 | 96.00 | 102.06 | 100.00 | 1.061 | 1.063 | 1.002 |
| 8 | 9530.78 | 121.90 | 105.30 | 110.14 | 10.68 | -4.39 | 110.68 | 95.61 | 94.50 | 102.87 | 100.00 | 1.158 | 1.089 | 0.940 |
| 9 | 9155.27 | 113.49 | 107.42 | 107.95 | 5.14 | -0.48 | 105.14 | 99.52 | 95.58 | 102.29 | 100.00 | 1.057 | 1.070 | 1.013 |
| 10 | 8052.59 | 109.32 | 99.37 | 101.24 | 7.98 | -1.84 | 107.98 | 98.16 | 94.34 | 102.95 | 100.00 | 1.100 | 1.091 | 0.992 |
| 11 | 8845.33 | 121.63 | 97.66 | 106.10 | 14.63 | -7.96 | 114.63 | 92.04 | 94.78 | 102.72 | 100.00 | 1.245 | 1.084 | 0.870 |
| 12 | 7885.70 | 108.36 | 97.66 | 100.18 | 8.16 | -2.52 | 108.16 | 97.48 | 94.84 | 102.68 | 100.00 | 1.110 | 1.083 | 0.976 |
| 13 | 6854.53 | 117.24 | 83.01 | 93.40 | 25.52 | -11.13 | 125.52 | 88.87 | 89.64 | 105.62 | 100.00 | 1.412 | 1.178 | 0.834 |
| 14 | 7212.16 | 112.41 | 83.90 | 95.81 | 17.33 | -12.43 | 117.33 | 87.57 | 97.33 | 101.36 | 100.00 | 1.340 | 1.041 | 0.777 |
| 15 | 7516.15 | 105.63 | 96.68 | 97.81 | 8.00 | -1.16 | 108.00 | 98.84 | 93.68 | 103.32 | 100.00 | 1.093 | 1.103 | 1.009 |
| 16 | 9936.09 | 119.73 | 112.73 | 112.45 | 6.47 | 0.25 | 106.47 | 100.25 | 93.69 | 103.31 | 100.00 | 1.062 | 1.103 | 1.038 |
| 17 | 8779.76 | 113.89 | 102.54 | 105.71 | 7.74 | -3.00 | 107.74 | 97.00 | 95.69 | 102.23 | 100.00 | 1.111 | 1.068 | 0.962 |
| 18 | 9638.07 | 116.70 | 108.76 | 110.75 | 5.37 | -1.80 | 105.37 | 98.20 | 96.64 | 101.72 | 100.00 | 1.073 | 1.053 | 0.981 |
| 19 | 8529.43 | 111.83 | 102.54 | 104.19 | 7.33 | -1.58 | 107.33 | 98.42 | 94.67 | 102.78 | 100.00 | 1.091 | 1.086 | 0.995 |
| 20 | 10526.18 | 129.42 | 109.55 | 115.75 | 11.81 | -5.35 | 111.81 | 94.65 | 94.49 | 102.87 | 100.00 | 1.181 | 1.089 | 0.922 |
| 21 | 8755.92 | 111.83 | 104.98 | 105.56 | 5.93 | -0.55 | 105.93 | 99.45 | 94.93 | 102.64 | 100.00 | 1.065 | 1.081 | 1.015 |
| 22 | 12809.04 | 158.24 | 108.12 | 127.68 | 23.93 | -15.32 | 123.93 | 84.68 | 95.29 | 102.44 | 100.00 | 1.464 | 1.075 | 0.735 |
| 23 | 8678.44 | 111.19 | 104.31 | 105.10 | 5.79 | -0.75 | 105.79 | 99.25 | 95.23 | 102.47 | 100.00 | 1.066 | 1.076 | 1.009 |
| 24 | 8410.22 | 111.19 | 101.67 | 103.46 | 7.47 | -1.73 | 107.47 | 98.27 | 94.69 | 102.76 | 100.00 | 1.094 | 1.085 | 0.992 |
| 25 | 9280.44 | 116.83 | 107.03 | 108.68 | 7.50 | -1.52 | 107.50 | 98.48 | 94.46 | 102.89 | 100.00 | 1.092 | 1.089 | 0.998 |
| 26 | 9506.94 | 114.77 | 107.42 | 110.00 | 4.34 | -2.34 | 104.34 | 97.66 | 98.14 | 100.94 | 100.00 | 1.068 | 1.029 | 0.963 |
| 27 | 9167.19 | 115.06 | 108.08 | 108.02 | 6.52 | 0.06 | 106.52 | 100.06 | 93.82 | 103.24 | 100.00 | 1.065 | 1.100 | 1.034 |
| 28 | 10216.24 | 123.77 | 109.86 | 114.03 | 8.54 | -3.65 | 108.54 | 96.35 | 95.62 | 102.26 | 100.00 | 1.127 | 1.069 | 0.949 |
| 29 | 9167.19 | 115.76 | 105.31 | 108.02 | 7.17 | -2.51 | 107.17 | 97.49 | 95.71 | 102.21 | 100.00 | 1.099 | 1.068 | 0.972 |
| 30 | 10496.38 | 123.16 | 109.86 | 115.58 | 6.56 | -4.95 | 106.56 | 95.05 | 98.73 | 100.64 | 100.00 | 1.121 | 1.019 | 0.909 |
| 31 | 11271.24 | 140.88 | 100.31 | 119.77 | 17.63 | -16.25 | 117.63 | 83.75 | 101.51 | 99.25 | 100.00 | 1.405 | 0.978 | 0.696 |
| 32 | 9614.23 | 120.62 | 104.83 | 110.62 | 9.04 | -5.23 | 109.04 | 94.77 | 96.77 | 101.66 | 100.00 | 1.151 | 1.051 | 0.913 |
| 33 | 9781.12 | 120.60 | 110.07 | 111.57 | 8.09 | -1.35 | 108.09 | 98.65 | 93.78 | 103.26 | 100.00 | 1.096 | 1.101 | 1.005 |
| 34 | 9787.08 | 118.90 | 108.96 | 111.61 | 6.54 | -2.38 | 106.54 | 97.62 | 96.15 | 101.98 | 100.00 | 1.091 | 1.061 | 0.972 |
| 35 | 9530.78 | 116.91 | 109.86 | 110.14 | 6.15 | -0.25 | 106.15 | 99.75 | 94.44 | 102.90 | 100.00 | 1.064 | 1.090 | 1.024 |
| 36 | 9852.65 | 123.16 | 102.54 | 111.98 | 9.99 | -8.43 | 109.99 | 91.57 | 99.29 | 100.36 | 100.00 | 1.201 | 1.011 | 0.841 |
| 37 | 8386.37 | 114.93 | 95.22 | 103.31 | 11.24 | -7.84 | 111.24 | 92.16 | 97.54 | 101.25 | 100.00 | 1.207 | 1.038 | 0.860 |
| 38 | 9787.08 | 124.32 | 103.44 | 111.61 | 11.39 | -7.32 | 111.39 | 92.68 | 96.86 | 101.61 | 100.00 | 1.202 | 1.049 | 0.873 |
| 39 | 7271.77 | 103.00 | 95.22 | 96.20 | 7.07 | -1.03 | 107.07 | 98.97 | 94.37 | 102.94 | 100.00 | 1.082 | 1.091 | 1.008 |
| 40 | 8821.49 | 113.26 | 106.55 | 105.96 | 6.89 | 0.56 | 106.89 | 100.56 | 93.03 | 103.68 | 100.00 | 1.063 | 1.114 | 1.048 |
| 41 | 9083.75 | 114.36 | 103.63 | 107.52 | 6.36 | -3.62 | 106.36 | 96.38 | 97.55 | 101.25 | 100.00 | 1.103 | 1.038 | 0.941 |
| 42 | 7551.91 | 104.30 | 95.22 | 98.04 | 6.38 | -2.88 | 106.38 | 97.12 | 96.79 | 101.65 | 100.00 | 1.095 | 1.050 | 0.959 |
| 43 | 7551.91 | 105.32 | 97.09 | 98.04 | 7.43 | -0.97 | 107.43 | 99.03 | 93.99 | 103.15 | 100.00 | 1.085 | 1.097 | 1.012 |
| 44 | 7230.04 | 101.28 | 94.95 | 95.93 | 5.58 | -1.02 | 105.58 | 98.98 | 95.69 | 102.23 | 100.00 | 1.067 | 1.068 | 1.002 |
| 45 | 9167.19 | 115.47 | 104.84 | 108.02 | 6.90 | -2.94 | 106.90 | 97.06 | 96.38 | 101.86 | 100.00 | 1.101 | 1.057 | 0.960 |
| 46 | 8451.94 | 110.70 | 101.61 | 103.72 | 6.73 | -2.03 | 106.73 | 97.97 | 95.63 | 102.26 | 100.00 | 1.089 | 1.069 | 0.981 |
| 47 | 10454.66 | 126.98 | 110.49 | 115.35 | 10.08 | -4.22 | 110.08 | 95.78 | 94.85 | 102.68 | 100.00 | 1.149 | 1.083 | 0.942 |
| 48 | 8726.12 | 111.56 | 103.42 | 105.38 | 5.86 | -1.86 | 105.86 | 98.14 | 96.26 | 101.92 | 100.00 | 1.079 | 1.059 | 0.982 |
| 49 | 7319.45 | 101.55 | 95.22 | 96.52 | 5.21 | -1.35 | 105.21 | 98.65 | 96.35 | 101.88 | 100.00 | 1.066 | 1.057 | 0.991 |
| 50 | 7247.93 | 119.13 | 87.40 | 96.04 | 24.04 | -9.00 | 124.04 | 91.00 | 88.59 | 106.24 | 100.00 | 1.363 | 1.199 | 0.880 |
| 51 | 9912.25 | 123.16 | 108.71 | 112.32 | 9.66 | -3.21 | 109.66 | 96.79 | 94.22 | 103.02 | 100.00 | 1.133 | 1.093 | 0.965 |
| 52 | 10216.24 | 125.85 | 108.37 | 114.03 | 10.36 | -4.97 | 110.36 | 95.03 | 95.35 | 102.41 | 100.00 | 1.161 | 1.074 | 0.925 |
| 53 | 7992.98 | 113.60 | 92.08 | 100.86 | 12.63 | -8.70 | 112.63 | 91.30 | 97.25 | 101.40 | 100.00 | 1.234 | 1.043 | 0.845 |
| 54 | 8380.41 | 111.83 | 97.20 | 103.28 | 8.28 | -5.88 | 108.28 | 94.12 | 98.13 | 100.95 | 100.00 | 1.150 | 1.029 | 0.894 |
| 55 | 8940.70 | 118.53 | 102.66 | 106.67 | 11.11 | -3.76 | 111.11 | 96.24 | 93.52 | 103.41 | 100.00 | 1.155 | 1.106 | 0.958 |
| 56 | 9816.89 | 119.33 | 109.86 | 111.78 | 6.76 | -1.71 | 106.76 | 98.29 | 95.30 | 102.43 | 100.00 | 1.086 | 1.075 | 0.990 |
| 57 | 8249.28 | 108.88 | 102.37 | 102.47 | 6.26 | -0.09 | 106.26 | 99.91 | 94.19 | 103.04 | 100.00 | 1.064 | 1.094 | 1.028 |
| 58 | 8356.57 | 110.70 | 101.55 | 103.13 | 7.34 | -1.53 | 107.34 | 98.47 | 94.61 | 102.81 | 100.00 | 1.090 | 1.087 | 0.997 |
| 59 | 7104.87 | 101.16 | 95.22 | 95.09 | 6.38 | 0.13 | 106.38 | 100.13 | 93.88 | 103.21 | 100.00 | 1.062 | 1.099 | 1.035 |
| 60 | 10031.46 | 121.51 | 108.87 | 112.99 | 7.54 | -3.65 | 107.54 | 96.35 | 96.51 | 101.79 | 100.00 | 1.116 | 1.055 | 0.945 |
| | | | | | | | | | | | | | | |
| | | | | | | | Average -> | | 95.44 | 102.38 | 100.00 | 1.137 | 1.073 | 0.951 |
| | | | | | | | St.Dev. -> | | 2.06 | 1.11 | 0.00 | 0.099 | 0.035 | 0.080 |

**Table S2.** Spherical equivalent volume diameters (Dxyz) calculated by combining SEM (Dxy) and AFM (Dz) reported diameters of the Au nanoparticles extracted from Fig. 15 in Ref. [3].

| S.No. | AFM(Dz) nm | SEM(Dxy) nm | Dxyz nm |
|---|---|---|---|
| 1 | 43.50 | 42.50 | 42.83 |
| 2 | 42.20 | 45.50 | 44.37 |
| 3 | 43.20 | 49.00 | 46.98 |
| 4 | 46.50 | 49.00 | 48.15 |
| 5 | 49.60 | 51.00 | 50.53 |
| 6 | 48.80 | 51.50 | 50.58 |
| 7 | 47.90 | 52.00 | 50.60 |
| 8 | 50.70 | 52.10 | 51.63 |
| 9 | 46.90 | 52.40 | 50.50 |
| 10 | 50.40 | 52.50 | 51.79 |
| 11 | 53.80 | 52.50 | 52.93 |
| 12 | 51.80 | 53.00 | 52.60 |
| 13 | 46.50 | 53.30 | 50.93 |
| 14 | 43.80 | 53.50 | 50.05 |
| 15 | 46.80 | 54.00 | 51.48 |
| 16 | 46.20 | 54.20 | 51.39 |
| 17 | 50.70 | 54.30 | 53.07 |
| 18 | 49.00 | 54.80 | 52.79 |
| 19 | 54.10 | 54.80 | 54.57 |
| 20 | 50.00 | 55.10 | 53.34 |
| 21 | 54.30 | 55.10 | 54.83 |
| 22 | 46.50 | 55.50 | 52.32 |
| 23 | 49.20 | 55.50 | 53.32 |
| 24 | 49.40 | 55.50 | 53.39 |
| 25 | 50.90 | 55.70 | 54.05 |
| 26 | 49.90 | 55.80 | 53.76 |
| 27 | 45.20 | 56.00 | 52.14 |
| 28 | 55.20 | 56.80 | 56.26 |
| 29 | 47.10 | 57.50 | 53.80 |
| 30 | 49.50 | 58.00 | 55.02 |
| 31 | 50.10 | 58.20 | 55.36 |
| 32 | 50.80 | 58.20 | 55.62 |
| 33 | 53.20 | 58.40 | 56.61 |
| 34 | 55.80 | 58.40 | 57.52 |
| 35 | 52.80 | 58.50 | 56.53 |
| 36 | 48.60 | 58.60 | 55.06 |
| 37 | 49.70 | 58.60 | 55.47 |
| 38 | 56.00 | 58.80 | 57.85 |
| 39 | 50.80 | 59.00 | 56.13 |
| 40 | 52.30 | 59.00 | 56.68 |

**Table S2 continued…**

| | | | |
|---|---|---|---|
| 41 | 43.80 | 59.50 | 53.72 |
| 42 | 49.70 | 59.50 | 56.04 |
| 43 | 46.90 | 59.80 | 55.15 |
| 44 | 45.80 | 60.00 | 54.83 |
| 45 | 51.10 | 60.50 | 57.19 |
| 46 | 46.00 | 60.70 | 55.34 |
| 47 | 51.10 | 61.20 | 57.63 |
| 48 | 50.70 | 61.30 | 57.54 |
| 49 | 58.30 | 61.40 | 60.35 |
| 50 | 52.10 | 61.70 | 58.32 |
| 51 | 49.00 | 62.30 | 57.51 |
| 52 | 53.40 | 62.40 | 59.24 |
| 53 | 50.20 | 62.90 | 58.34 |
| 54 | 52.30 | 63.30 | 59.40 |
| 55 | 54.10 | 63.40 | 60.13 |
| 56 | 54.60 | 63.80 | 60.57 |
| 57 | 52.70 | 64.00 | 59.99 |
| 58 | 45.40 | 64.50 | 57.38 |
| 59 | 39.90 | 64.50 | 54.96 |
| 60 | 50.10 | 64.60 | 59.35 |
| 61 | 53.50 | 64.70 | 60.73 |
| 62 | 54.40 | 64.90 | 61.19 |
| 63 | 36.50 | 65.00 | 53.63 |
| 64 | 48.80 | 65.30 | 59.26 |
| 65 | 51.10 | 66.00 | 60.60 |
| 66 | 52.60 | 67.80 | 62.30 |
| 67 | 59.60 | 68.00 | 65.08 |
| 68 | 47.90 | 68.40 | 60.74 |
| | | | |
| Average-> | 49.73 | 58.17 | 55.14 |
| St.Dev.-> | 4.12 | 5.39 | 4.13 |

**Table S3. Spherical equivalent volume diameters of the pebbles measured using the two independent methods.**

(a) Spherical equivalent volume diameters of the irregularly shaped pebbles calculated by combining the projection area and the peak-height. Images of irregularly shaped, aquarium glass pebbles spread onto a smooth surface were acquired and then analyzed using ImageJ to determine Aproj, from which Dxy was calculated. The peak-height of each pebble (Dz) was measured using a digital caliper. The volume was calculated using Dxy and Dz assuming an ellipsoidal shape. From this, an experimental spherical equivalent volume diameter (Dxyz1) was then calculated.

(b) Spherical equivalent volume diameters of the irregularly shaped pebbles calculated using mass and density. The mass of each pebble was measured, from which volume was calculated using the density of the pebbles (we measured the density of the pebbles to be $2.55 \times 10^{-3}$ g/mm$^{3}$). The spherical equivalent volume diameter (Dxyz2) was then calculated based on the volume.

(c) Shape ratios. The pebbles used have a mean R3 ratio of 1.1 indicating that they are dominated relatively more by flattening than by elongation.

Averages, standard deviations, and standard deviations of the mean (SDOM $= \frac{Std.Dev.}{n^{1/2}}$ , where n is the number of measurements) of Dxyz1 and Dxyz2 are provided at the bottom of the table. Statistical standard uncertainty ($\sigma = \sqrt{\sum SDOM^2}$) of Dxyz1 and Dxyz2 as estimated using error propagation (quadrature addition) is 0.172. The fact that the absolute difference (0.16) between Dxyz1 and Dxyz2 is less than $\sigma$ means that the difference is consistent with zero. This implies that Dxyz1 and Dxyz2 are statistically in agreement.

| | a | | | | | | | b | | | c | | |
|---|---|---|---|---|---|---|---|---|---|---|---|---|---|
| | Dxyz calculated using projection area and height | | | | | | | Dxyz calculated using mass and density | | | Shape ratios | | |
| S.No. | Aproj | Dx | Dy | Dxy | Dz | Volume | Dxyz1 | Mass | Volume | Dxyz2 | R1 | R2 | R3 |
| | $mm^2$ | mm | mm | mm | mm | $mm^3$ | mm | g | $mm^3$ | mm | Dx/Dy | Dxy/Dz | R2/R1 |
| 1 | 596.78 | 30.49 | 25.78 | 27.56 | 18.56 | 7384.17 | 24.16 | 20.43 | 8011.76 | 24.82 | 1.18 | 1.48 | 1.26 |
| 2 | 617.50 | 30.43 | 26.28 | 28.03 | 18.03 | 7422.39 | 24.20 | 20.68 | 8109.80 | 24.92 | 1.16 | 1.55 | 1.34 |
| 3 | 557.03 | 29.05 | 24.43 | 26.63 | 16.70 | 6201.59 | 22.79 | 16.23 | 6364.71 | 22.99 | 1.19 | 1.59 | 1.34 |
| 4 | 511.92 | 28.53 | 24.18 | 25.53 | 18.77 | 6405.81 | 23.04 | 16.23 | 6364.71 | 22.99 | 1.18 | 1.36 | 1.15 |
| 5 | 443.41 | 27.25 | 21.21 | 23.76 | 22.45 | 6636.37 | 23.31 | 16.39 | 6427.45 | 23.07 | 1.28 | 1.06 | 0.82 |
| 6 | 564.64 | 31.51 | 23.28 | 26.81 | 17.86 | 6722.99 | 23.41 | 17.48 | 6854.90 | 23.57 | 1.35 | 1.50 | 1.11 |
| 7 | 584.94 | 30.83 | 25.15 | 27.28 | 20.56 | 8017.58 | 24.83 | 21.00 | 8235.29 | 25.05 | 1.23 | 1.33 | 1.08 |
| 8 | 567.04 | 28.20 | 26.55 | 26.86 | 19.68 | 7439.53 | 24.22 | 18.60 | 7294.12 | 24.06 | 1.06 | 1.37 | 1.28 |
| 9 | 595.87 | 30.10 | 24.98 | 27.54 | 14.32 | 5688.52 | 22.14 | 16.75 | 6568.63 | 23.23 | 1.20 | 1.92 | 1.60 |
| 10 | 522.56 | 27.60 | 23.92 | 25.79 | 16.62 | 5789.99 | 22.28 | 16.35 | 6411.76 | 23.05 | 1.15 | 1.55 | 1.34 |
| 11 | 468.22 | 26.74 | 22.82 | 24.41 | 22.26 | 6948.38 | 23.67 | 17.60 | 6901.96 | 23.62 | 1.17 | 1.10 | 0.94 |
| 12 | 538.13 | 30.56 | 22.18 | 26.17 | 22.31 | 8003.80 | 24.81 | 21.48 | 8423.53 | 25.24 | 1.38 | 1.17 | 0.85 |
| 13 | 528.34 | 28.88 | 23.39 | 25.93 | 15.46 | 5445.44 | 21.82 | 14.75 | 5784.31 | 22.27 | 1.23 | 1.68 | 1.36 |
| 14 | 438.27 | 28.17 | 20.15 | 23.62 | 22.66 | 6620.72 | 23.29 | 17.45 | 6843.14 | 23.55 | 1.40 | 1.04 | 0.75 |
| 15 | 546.39 | 29.37 | 23.54 | 26.37 | 21.65 | 7886.17 | 24.69 | 20.53 | 8050.98 | 24.86 | 1.25 | 1.22 | 0.98 |
| 16 | 509.45 | 27.87 | 23.99 | 25.46 | 16.26 | 5522.47 | 21.93 | 14.42 | 5654.90 | 22.10 | 1.16 | 1.57 | 1.35 |
| 17 | 521.72 | 27.92 | 23.28 | 25.77 | 17.00 | 5912.79 | 22.43 | 15.88 | 6227.45 | 22.82 | 1.20 | 1.52 | 1.26 |
| 18 | 575.98 | 29.74 | 25.25 | 27.08 | 19.71 | 7568.38 | 24.36 | 18.75 | 7352.94 | 24.12 | 1.18 | 1.37 | 1.17 |
| 19 | 606.93 | 30.32 | 25.65 | 27.79 | 17.70 | 7161.79 | 23.91 | 19.63 | 7698.04 | 24.49 | 1.18 | 1.57 | 1.33 |
| 20 | 596.71 | 30.67 | 24.93 | 27.56 | 19.53 | 7769.18 | 24.57 | 20.52 | 8047.06 | 24.86 | 1.23 | 1.41 | 1.15 |
| 21 | 574.37 | 28.07 | 26.66 | 27.04 | 23.67 | 9063.53 | 25.86 | 21.10 | 8274.51 | 25.09 | 1.05 | 1.14 | 1.08 |
| 22 | 466.39 | 30.19 | 19.39 | 24.36 | 24.30 | 7555.47 | 24.34 | 20.14 | 7898.04 | 24.70 | 1.56 | 1.00 | 0.64 |
| 23 | 565.91 | 29.00 | 25.62 | 26.84 | 19.41 | 7322.88 | 24.09 | 20.96 | 8219.61 | 25.04 | 1.13 | 1.38 | 1.22 |
| 24 | 464.06 | 27.39 | 21.30 | 24.30 | 25.73 | 7960.19 | 24.77 | 19.96 | 7827.45 | 24.63 | 1.29 | 0.94 | 0.73 |
| 25 | 553.01 | 30.56 | 22.80 | 26.53 | 21.56 | 7948.61 | 24.76 | 19.71 | 7729.41 | 24.53 | 1.34 | 1.23 | 0.92 |
| 26 | 511.92 | 29.69 | 21.50 | 25.53 | 23.30 | 7951.81 | 24.76 | 20.38 | 7992.16 | 24.80 | 1.38 | 1.10 | 0.79 |
| 27 | 551.46 | 31.24 | 22.57 | 26.49 | 23.55 | 8657.94 | 25.47 | 19.88 | 7796.08 | 24.60 | 1.38 | 1.12 | 0.81 |
| 28 | 486.69 | 26.74 | 23.59 | 24.89 | 22.13 | 7180.26 | 23.93 | 18.12 | 7105.88 | 23.85 | 1.13 | 1.12 | 0.99 |
| 29 | 499.80 | 28.38 | 22.59 | 25.22 | 23.34 | 7776.83 | 24.58 | 18.99 | 7447.06 | 24.23 | 1.26 | 1.08 | 0.86 |
| 30 | 533.14 | 29.06 | 23.78 | 26.05 | 22.84 | 8117.87 | 24.93 | 20.13 | 7894.12 | 24.70 | 1.22 | 1.14 | 0.93 |
| 31 | 661.70 | 31.28 | 28.16 | 29.02 | 15.04 | 6634.61 | 23.31 | 17.56 | 6886.27 | 23.60 | 1.11 | 1.93 | 1.74 |
| 32 | 647.32 | 33.43 | 24.19 | 28.70 | 16.51 | 7124.81 | 23.87 | 20.74 | 8133.33 | 24.95 | 1.38 | 1.74 | 1.26 |
| 33 | 486.55 | 29.04 | 20.78 | 24.88 | 22.41 | 7269.00 | 24.03 | 17.38 | 6815.69 | 23.52 | 1.40 | 1.11 | 0.79 |
| 34 | 469.91 | 27.30 | 22.56 | 24.46 | 22.73 | 7120.73 | 23.87 | 18.15 | 7117.65 | 23.86 | 1.21 | 1.08 | 0.89 |
| 35 | 570.98 | 28.74 | 25.77 | 26.96 | 17.42 | 6631.03 | 23.31 | 18.31 | 7180.39 | 23.93 | 1.12 | 1.55 | 1.39 |
| 36 | 583.87 | 29.68 | 24.37 | 27.26 | 17.83 | 6940.32 | 23.66 | 18.99 | 7447.06 | 24.23 | 1.22 | 1.53 | 1.26 |
| 37 | 501.91 | 27.81 | 23.06 | 25.27 | 19.86 | 6645.30 | 23.32 | 17.87 | 7007.84 | 23.74 | 1.21 | 1.27 | 1.06 |
| 38 | 570.77 | 29.27 | 24.47 | 26.95 | 16.28 | 6194.79 | 22.78 | 16.85 | 6607.84 | 23.28 | 1.20 | 1.66 | 1.38 |
| 39 | 527.64 | 29.70 | 22.20 | 25.91 | 22.11 | 7777.37 | 24.58 | 19.91 | 7807.84 | 24.61 | 1.34 | 1.17 | 0.88 |
| 40 | 619.41 | 33.05 | 25.30 | 28.08 | 16.66 | 6879.54 | 23.59 | 17.78 | 6972.55 | 23.70 | 1.31 | 1.69 | 1.29 |
| 41 | 595.30 | 29.84 | 25.72 | 27.53 | 17.89 | 7099.96 | 23.84 | 19.36 | 7592.16 | 24.38 | 1.16 | 1.54 | 1.33 |
| 42 | 508.60 | 28.32 | 22.47 | 25.44 | 15.20 | 5153.79 | 21.43 | 14.68 | 5756.86 | 22.23 | 1.26 | 1.67 | 1.33 |
| 43 | 440.59 | 26.64 | 21.59 | 23.68 | 24.60 | 7225.68 | 23.98 | 17.72 | 6949.02 | 23.67 | 1.23 | 0.96 | 0.78 |
| 44 | 553.36 | 28.31 | 24.41 | 26.54 | 18.52 | 6832.20 | 23.54 | 17.71 | 6945.10 | 23.67 | 1.16 | 1.43 | 1.24 |
| 45 | 687.92 | 30.89 | 28.04 | 29.59 | 16.85 | 7727.59 | 24.53 | 20.07 | 7870.59 | 24.68 | 1.10 | 1.76 | 1.59 |
| 46 | 563.37 | 29.58 | 24.78 | 26.78 | 15.06 | 5656.25 | 22.10 | 14.98 | 5874.51 | 22.38 | 1.19 | 1.78 | 1.49 |
| 47 | 600.31 | 31.02 | 24.40 | 27.64 | 20.48 | 8196.16 | 25.01 | 20.81 | 8160.78 | 24.98 | 1.27 | 1.35 | 1.06 |
| 48 | 463.00 | 29.26 | 18.66 | 24.27 | 21.65 | 6682.58 | 23.37 | 17.47 | 6850.98 | 23.56 | 1.57 | 1.12 | 0.72 |
| 49 | 504.38 | 29.65 | 22.21 | 25.34 | 23.27 | 7824.58 | 24.63 | 19.12 | 7498.04 | 24.28 | 1.34 | 1.09 | 0.82 |
| 50 | 508.47 | 29.85 | 21.66 | 25.44 | 20.62 | 6989.71 | 23.72 | 17.40 | 6823.53 | 23.53 | 1.38 | 1.23 | 0.90 |
| 51 | 582.54 | 30.93 | 23.81 | 27.23 | 21.20 | 8233.29 | 25.05 | 19.18 | 7521.57 | 24.31 | 1.30 | 1.28 | 0.99 |
| 52 | 615.46 | 31.88 | 24.30 | 27.99 | 16.66 | 6835.70 | 23.54 | 19.33 | 7580.39 | 24.37 | 1.31 | 1.68 | 1.28 |
| 53 | 495.14 | 27.83 | 23.44 | 25.10 | 22.21 | 7331.43 | 24.10 | 17.50 | 6862.75 | 23.57 | 1.19 | 1.13 | 0.95 |
| 54 | 403.09 | 26.08 | 20.71 | 22.65 | 23.50 | 6315.12 | 22.93 | 16.32 | 6400.00 | 23.03 | 1.26 | 0.96 | 0.77 |
| | | | | | | | | | | | | | |
| | | | | | | Average -> | **23.80** | | | **23.96** | **1.25** | **1.36** | **1.10** |
| | | | | | | Standard deviation -> | 0.97 | | | 0.82 | 0.11 | 0.26 | 0.26 |
| | | | | | | SDOM -> | 0.13 | | | 0.11 | | | |
| | | | | | | | | σ = | 0.17 | | | | |
| | | | | | | | | $\lvert Dxyz1 - Dxyz2 \rvert$ = | 0.16 | | | | |